\documentclass{kluwer}    
\newdisplay{guess}{Conjecture}
\usepackage{graphicx}
\newcommand{\ben}{\begin{enumerate}}
\newcommand{\een}{\end{enumerate}}
\newcommand{\bfig}{\begin{figure}}
\newcommand{\efig}{\end{figure}}
\newcommand{\beq}{\begin{equation}}
\newcommand{\eeq}{\end{equation}}
\newcommand{\mbf}{\mathbf}
\begin{document}          
\begin{article}
\begin{opening}         
\title{Turbulence in the Solar Atmosphere: Manifestations and 
Diagnostics via Solar Image Processing\thanks{Based on the author's 
contributed talk ``Manifestations and Diagnostics of Turbulence in 
the Solar Atmosphere'', presented at the Solar Image Processing 
Workshop II, Annapolis, Maryland, USA, 3 - 5 November 2004}} 
\author{Manolis K. \surname{Georgoulis}}  
\runningauthor{Manolis Georgoulis}
\runningtitle{Turbulence: A Physical Context for Solar Image Processing}
\institute{The Johns Hopkins University Applied Physics Laboratory, 
11100 Johns Hopkins Rd., Laurel, MD 20723, USA}
\date{January 31, 2005}
\begin{abstract}
Intermittent magnetohydrodynamical turbulence is most likely at work in the 
magnetized solar atmosphere. As a result, an array of scaling and multi-scaling 
image processing techniques can be used to measure the expected 
self-organization of solar magnetic fields. While these techniques 
advance our understanding of the physical system at work, it is unclear whether 
they can be used to predict solar eruptions, thus 
obtaining a practical significance for space weather. We address part of 
this problem by focusing on solar active regions and by investigating the usefulness 
of scaling and multi-scaling image processing techniques in solar flare prediction. 
Since solar flares exhibit spatial and temporal intermittency, we suggest that 
they are the products of instabilities subject to a critical 
threshold in a turbulent magnetic configuration. 
The identification of this threshold in scaling and multi-scaling spectra 
would then contribute meaningfully to the prediction of solar flares. 
We find that the fractal dimension of solar magnetic fields 
and their multifractal spectrum of generalized correlation dimensions do not have 
significant predictive ability. The respective 
multifractal structure functions and their inertial-range scaling exponents, 
however, probably provide some statistical distinguishing 
features between flaring and non-flaring active regions. More importantly, 
the temporal 
evolution of the above scaling exponents in flaring active regions probably shows 
a distinct behavior starting a few hours prior to a flare and therefore 
this temporal behavior may be practically useful in flare prediction. The results of 
this study need to be validated by more comprehensive works over a large number of 
solar active regions. Sufficient statistics may also establish 
critical thresholds in the values of the multifractal structure functions and/or 
their scaling exponents above which a flare may be predicted with a high 
level of confidence.  
\end{abstract}
\end{opening}           
\section{Introduction}  
The magnetized solar atmosphere can be viewed both locally (solar active 
regions) and globally as an externally driven, dissipative, nonlinear 
dynamical system. The evolution of the system is largely dictated by the 
configuration of the magnetic field vector, which is subject to 
boundary-induced perturbations. These perturbations are either the emergence 
of additional magnetic flux from the solar interior or the shuffling caused 
by horizontal motions on the lower atmospheric boundary, 
the solar photosphere. Flux emergence and boundary flows, therefore, play the 
role of the external driver. The response of the system to external perturbations 
is nonlinear, consists of multiple manifestations,  
and occurs after a variable time delay with respect to the 
driving perturbation. This interval is necessary for the onset of acoustic, 
Alfv\'{e}n, or magnetosonic, waves, and for the accumulation of electric currents 
in multiple locations within the magnetic configuration. The qualitative, 
first-order, description of the system is complete when dissipative processes 
are considered. These processes dissipate part of the non-potential magnetic 
energy stored in the magnetic structure in the form of electric currents. 
Currents are flowing both along and perpendicular to the magnetic field lines 
because these lines are twisted and clumped in braids, 
while localized shear between areas of different connectivity, either unipolar or 
multi-polar, creates additional current interfaces in the non-potential 
configuration. Dissipative events are abundant and can be intermittent, 
such as eruptions (flares and corona ejecta), or nearly continuous, such 
as the (homogeneous or inhomogeneous) heating processes in the solar chromosphere 
and corona. Although the energy content of the above processes is known 
and comes from the reservoir of free magnetic energy, solid evidence of their 
triggering via magnetic reconnection or wave damping and knowledge of their 
occurrence frequency are profoundly lacking. 

Other key factors that preclude further understanding of solar magnetic fields 
are their {\it filamentary} nature (Parker 1979) and an ever-acting  
{\it competition} between two intrinsic tendencies: clustering and 
fragmentation of elemental magnetic flux tubes in the solar atmosphere 
(see, e.g., Abramenko and Longcope 2005). It appears that magnetic fragmentation 
is the prevailing tendency, with clustering dominating temporarily and locally 
during the formation of 
solar active regions. These entities occupy a small 
fraction of the photospheric surface at low latitudes, while the rest is occupied 
by small-scale magnetic flux elements. 

To realize the physics of energy dissipation in the solar atmosphere 
the concept of {\it turbulence} has been proved a cornerstone in our 
understanding of solar magnetic fields. Most proposed 
solar dynamo mechanisms dealing with 
the generation of solar magnetic fields deep in the convection zone and the 
buoyant rise of magnetic flux tubes toward the atmosphere 
involve Kolmogorov's theory of fluid turbulence 
(Cattaneo, Emonet, and Weiss 2003; Archontis, Dorch, and Nordlund 2003 and 
references therein). There is a sizable amount of 
evidence that turbulence is extended in the 
magnetically dominated solar atmosphere in the form of magnetohydrodynamic (MHD) 
turbulence. In essence, a turbulent evolution requires (1) 
a few large-scale, coherent, current-carrying magnetic structures, (2) ideal 
(non-dissipative) fragmentation of the free magnetic energy within an inertial 
range of length scales, and (3) a critical length scale below which magnetic 
resistivity sets in and releases part of the fragmented free energy. All of the 
above requirements are satisfied in modeled active regions and are supported 
by observations of actual solar active regions: 
the vector potential and the photospheric flows are organized in a few large-scale 
structures, while the electric current density and hence the magnetic free 
energy are distributed within numerous small-scale structures with linear sizes 
extending down to our present observational limit 
(e.g., Einaudi {\it et al.,} 1996; Georgoulis, Velli, and Einaudi 1998; 
Chae 2001; Dmitruk {\it et al.,} 2002). The first and the 
second process are known as an inverse and a direct cascade, respectively. 
Dissipation occurs when the length scale of the current structures becomes 
comparable to the Taylor microscale ($\sim 3 \times 10^3\;cm$; 
Biskamp and Welter 1989), or even smaller 
($\sim 20\;cm$) according to Kraichnan's (1965) interpretation of a homogeneous 
and stationary MHD turbulence. Regardless of whether the above numbers are 
correct, the necessity of small spatial scales in order to achieve magnetic 
reconnection and subsequent current dissipation 
is most likely the reason why we lack evidence of flare and sub-flare triggering. 

The phenomenology of turbulence in the solar atmosphere suggests a hierarchical 
{\it self-organization} of the physical parameters of the system. By 
self-organization we mean the reduction of the many 
parameters (degrees of freedom) exhibited by the complex 
solar magnetic structures to a small number of {\it significant} degrees of 
freedom that dictate the system's evolution (see, e.g., Nicolis and Prigogine 
1989). Assuming nonlinear 
self-organization directly contributes a number of powerful tools borrowed from 
the theory of nonlinear dynamical systems that can help assess the 
evolution of solar magnetic fields without necessarily elaborating on the 
detailed physics of the system. Such techniques are the percolation theory 
(Stauffer and Aharony 1994) 
and the concept of Self-Organized Criticality 
(SOC; Bak, Tang, and Wiesenfield 1987; Bak 1996 for a review) which have been 
successfully applied to solar physics. Percolation achieves self-organization 
via a competition of probabilities that regulate the clustering, fragmentation, 
and diffusion of solar magnetic fields and has reproduced the magnetic flux 
emergence and the formation of active regions in the Sun (Wentzel and Seiden 1992; 
Seiden and Wentzel 1996; Vlahos {\it et al.,} 2002). SOC achieves 
self-organization without the finely tuned competing probabilities but using a 
critical threshold of a certain parameter 
(magnetic discontinuities, electric current density, etc.) 
whose excession, subject to external forcing, 
leads to an instability. According to local situation at the vicinity of 
the instability, a cascade of similar instabilities may occur 
as a result of the initial event as a domino-effect- or an avalanche-type process. 
SOC has described the triggering of solar flares and has provided a distinction 
between flares and sub-flares based on the size of the resulting avalanches 
(Lu and Hamilton 1991; Lu {\it et al.,} 1993; Vlahos {\it et al.,} 1995; 
Georgoulis and Vlahos 1996; 1998). 
Large avalanches correspond to flares, while small avalanches or nearly single 
events correspond to microflares or nanoflares, respectively. Notice, 
nonetheless, that the analogy between turbulence and SOC is imperfect 
in that the creation of an avalanche of elementary instabilities is yet to be 
definitively demonstrated in a turbulent simulation 
(Charbonneau {\it et al.,} 2001). Reversely, it has been shown that the large, 
well-observed, solar flares can only be achieved by means of a SOC-type avalanche 
process which relaxes a large number of localized elementary instabilities 
(Vlahos and Georgoulis 2004). 

A necessary but not sufficient condition for self-organization is the spatial 
and temporal {\it self-similarity} in the studied system. Self-similarity 
means that the system's behavior does not depend on the temporal and spatial 
scales present: a magnification of an active-region plage, 
for example, reveals additional complex structure in small scales provided that 
the spatial resolution of the observing instrument is sufficient. Spatial 
self-similarity is a consequence of the filamentary nature of solar magnetic 
fields. In addition, the magnification of the temporal X-ray profile of a flare 
reveals additional temporal structure of the emission. If a critical threshold 
is involved in a self-similar (fractal) process, then spatial and temporal 
{\it intermittency} are also expected: intense 
X-ray flare emission is obtained 
within a very short interval compared to typical time scales of the evolution 
in the source 
active region, while the flaring volume is small compared to the volume of 
the active region regardless of the flare size. Intermittency and self-similarity 
can be measured and monitored via an array of image processing techniques, 
such as multi-scaling analysis of fractal and multi-fractal structures, 
image contrast enhancement, and pattern recognition. As diverse as they 
may appear, these techniques help understand and quantify the 
turbulence and intermittency present  
in the solar atmosphere. Therefore, they help understand the system at work in a 
way that complements the traditional approaches of solar physics. Numerous models 
and observational works have revealed the fractality and multi-fractality of  
active regions and their components (sunspots, plages), of small-scale magnetic 
concentrations in the ``Quiet'' Sun, and of the white-light magneto-convection 
(granulation and supergranulation) patterns (e.g., Roudier and Muller 1987; 
Lawrence 1991; Schrijver {\it et al.,} 1992; Jan{\ss}en, V\"{o}gler, and 
Kneer 2003). These studies have 
updated our knowledge of solar magnetic fields by showcasing the 
tremendous complexity exhibited by the processes at work. We have also learned that  
a possibly self-organized system can be very complex in its response, despite the 
few degrees of freedom that dictate this response. Despite these advances, however, 
the contribution of the various image processing techniques in understanding 
specific properties of the system with practical interest is controversial and 
far from clear: for example, although we can detect the buildup of magnetic flux 
and electric currents in an active region via image processing, 
we cannot predict when, where, or 
whether a major eruption will occur. Several conflicting accounts appear in the 
literature but, to our best knowledge, the predictive ability of image processing 
techniques has not been comprehensively investigated yet. 

In this paper we acknowledge this problem and we suggest ways to tackle it 
by focusing on the multi-scaling properties of solar magnetic fields. 
Applications with practical space weather interest include flare and 
coronal mass ejection (CME) prediction. We will hereby focus on solar active 
regions (ARs) so we focus on flare, rather than CME, prediction, as the 
relationship between CMEs and the ``source'' ARs is unclear. 
In view of the observed intermittency of the solar flare phenomenon, 
our suggestion is to uncover likely 
{\it critical thresholds} of fractal and multifractal parameters 
whose excession might cause the triggering of a flare. We investigate whether 
this can be achieved by the existing array of image processing techniques and 
having the turbulent physics of the system in mind. 
In Section 2 we review some of the existing scaling and multi-scaling image 
processing techniques. In Section 3 
we illustrate the possible significance of these techniques in flare prediction.  
In Section 4 we summarize the discussion and present our conclusions. 
\section{Understanding and Quantifying Turbulence via Image Processing} 
\subsection{Flaring and Subflaring Activity}
Although much of the image processing applied to solar physics has been 
focused on solar magnetic fields (Section 2.2), numerous studies have 
revealed the intermittent and self-similar nature of the energy release 
process in actual dissipative structures. 
Pattern recognition of energy release events has been applied to 
active-region X-ray bright points and transient brightenings 
(e.g., Shimizu 1995), extreme ultraviolet (EUV) bright points 
(Aletti {\it et al.,} 2000; Benz and Krucker 2002; Aschwanden and Parnell 2002 
and references therein), and off-band $H\alpha$ bright points, otherwise known 
as Ellerman bombs (EBs; Georgoulis {\it et al.,} 2002). To our knowledge, 
such analyses 
have not been applied to full-scale flares observed in X-rays on in EUV 
wavelengths and we believe that a such study is long overdue. The identification 
of the emitting areas in the above works allows statistical studies in terms 
of the events' area, duration, intensity and, in some cases, calculated energy 
content. Given a statistically sufficient number of events, one may construct 
the probability distribution function (PDF) for each of the above parameters. 
These PDFs invariably show well-defined power laws which underline the 
self-similar nature of the energy release process. Intermittency is also 
evident since the total area occupied by the emitting structures at any given 
time is small compared to the observational field of view. Examples of pattern 
recognition in subflaring activity are given in Figure \ref{f1}. Although the 
existence of power laws is universal in flares and subflares, 
the value of the power-law index for the total-energy PDF for subflares is 
strongly debated. This value has implications 
on the heating efficiency of subflares in the solar atmosphere (see, e.g., 
Hudson 1991). In any case, the self-similar and intermittent nature of 
the energy release process causing the power-law PDFs appears beyond question. 
\bfig[t]
\centerline{\includegraphics[width=12.cm,height=10.cm,angle=0]{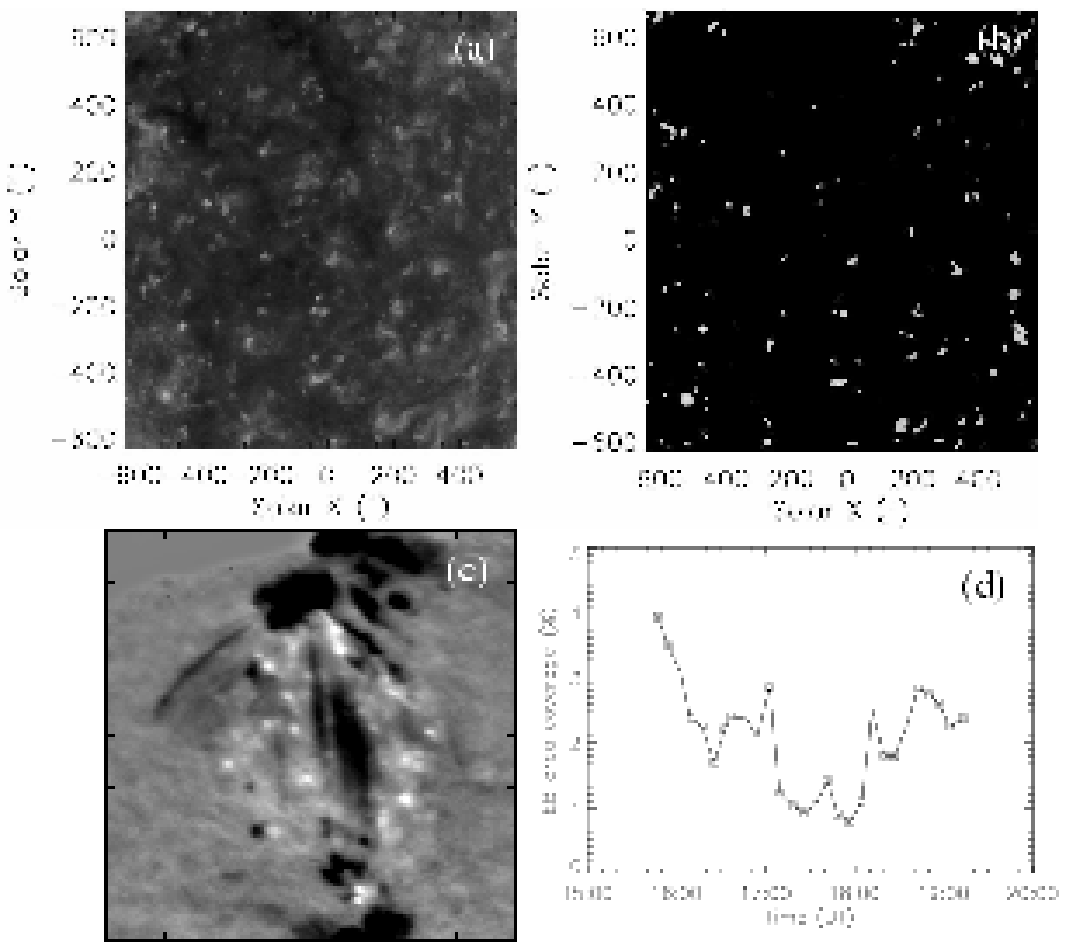}}
\caption{Pattern recognition in solar images of subflares: (a) Quiet Sun 
image from SoHO/EIT showing numerous EUV bright points, and (b) These 
bright points identified and singled out (from Alletti {\it et al.,} 2000). 
(c) Off-band $H\alpha$ image from the Flare Genesis Experiment (FGE) 
showing numerous Ellerman bombs (EBs). Tic mark separation is $10''$. 
(d) Total area coverage of EBs normalized over the total field of view area 
for the entire FGE observations on January 25, 2000. The small percentage of 
total area coverage illustrates the intermittency of the EB phenomenon.}    
\label{f1}
\efig

To quantify the degree of self-similarity in solar images such as those of 
Figure \ref{f1}, one may calculate the {\it fractal dimension} 
of the energy release patterns. A such study has been performed for EBs 
by Georgoulis {\it et al.,} (2002). They utilized a convenient box-counting method 
of measuring the fractal dimension (e.g., Mandelbrot 1983), as follows: 
Each off-band $H\alpha$ image (Figure \ref{f1}c) is covered by a two-dimensional 
uniform and rectangular grid consisting of boxes with linear size 
$\lambda$. If the field of view has a linear size $L$, then it 
is covered by $(L/\lambda)^2$ boxes with dimensionless area $\varepsilon ^2$, 
where $\varepsilon =(\lambda /L)$. Out of the $(L/\lambda)^2$ boxes, one then 
counts the number $N(\lambda)$ of boxes that contain at least part of an EB. 
Then, by varying $\lambda$, or, equivalently, 
the dimensionless area $\varepsilon$ of each box, one may study the variability 
of $N(\lambda)$ vs. $\lambda$ or $\varepsilon$. One obtains 
\beq
N(\varepsilon) \propto ({{1} \over {\varepsilon}})^{\mathcal{D}}\;\;,
\label{frd}
\eeq
where $\mathcal{D}$ is the (box-counting) fractal dimension of EBs. Other 
definitions of the fractal dimension can also be found in the literature, but 
adopting any one of them is sufficient to show whether a studied structure is 
fractal or not. The value of $\mathcal{D}$ can be at most equal to the Euclidean 
dimension of the image space for non-fractal structures. In our two-dimensional 
images, $\mathcal{D} =2$ would imply a spatially homogeneous, non-fractal, 
event distribution, 
while $\mathcal{D} <2$ would imply a fractal structure. The smaller the fractal 
dimension, the stronger the fragmenting tendency of the structures. 
For $\mathcal{D} <1$ in a two-dimensional image one obtains ``dust fractals'' 
with a dominant fragmentation tendency that precludes any mid- or large-scale 
structure clustering. 

\bfig[t]
\centerline{\includegraphics[width=4.cm,height=3.cm,angle=0]{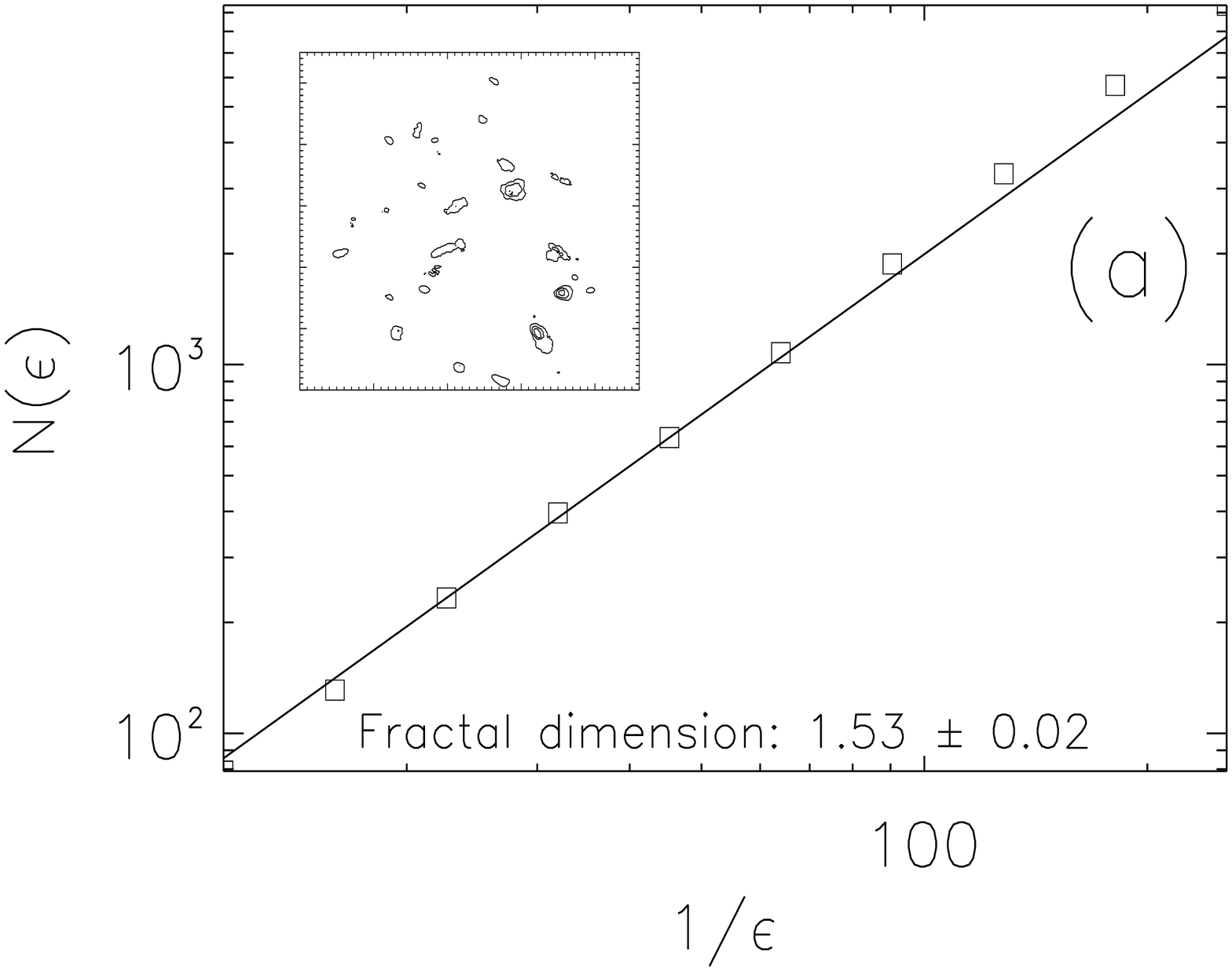}
            \includegraphics[width=4.cm,height=3.cm,angle=0]{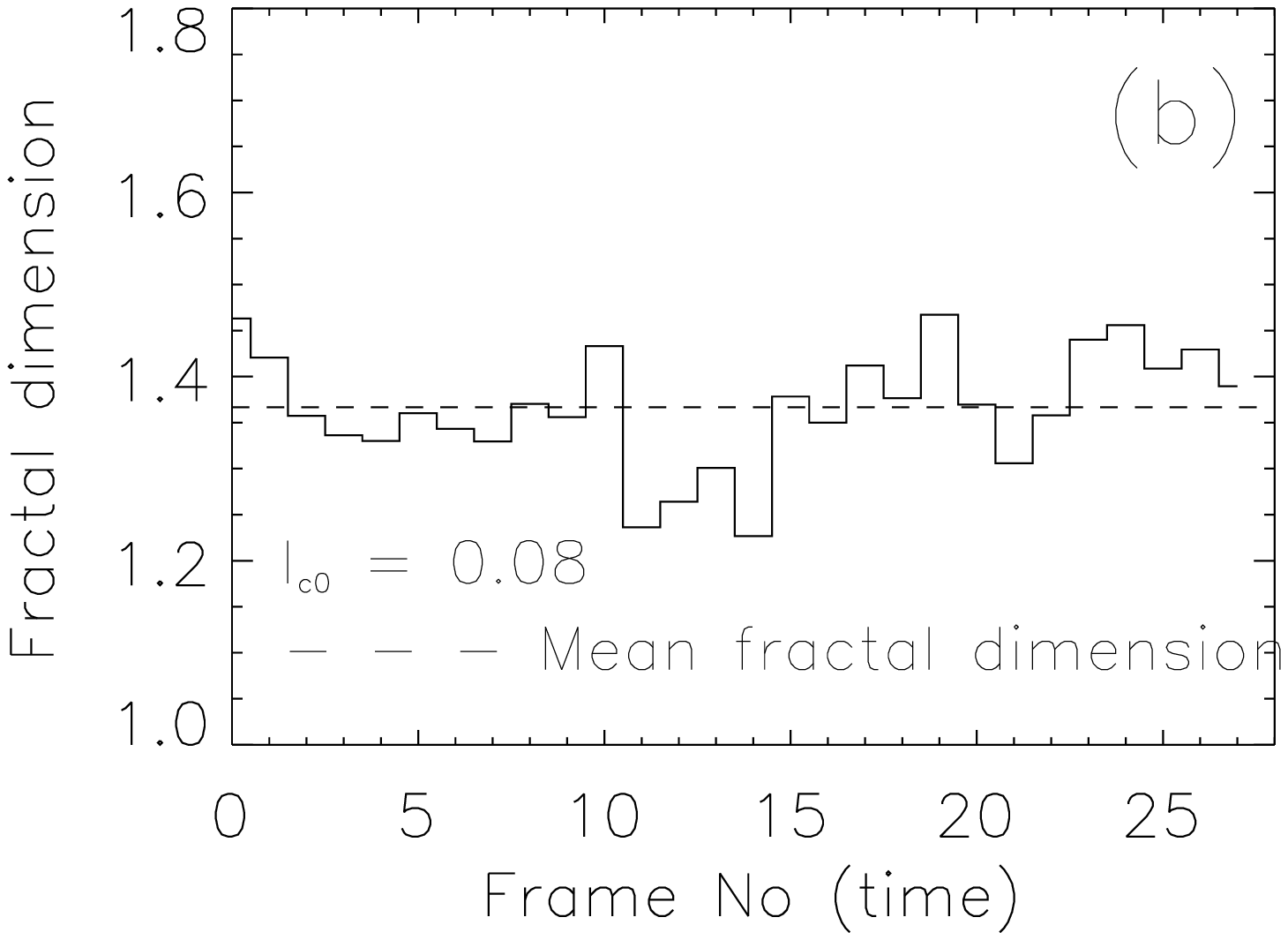}
            \includegraphics[width=4.cm,height=3.cm,angle=0]{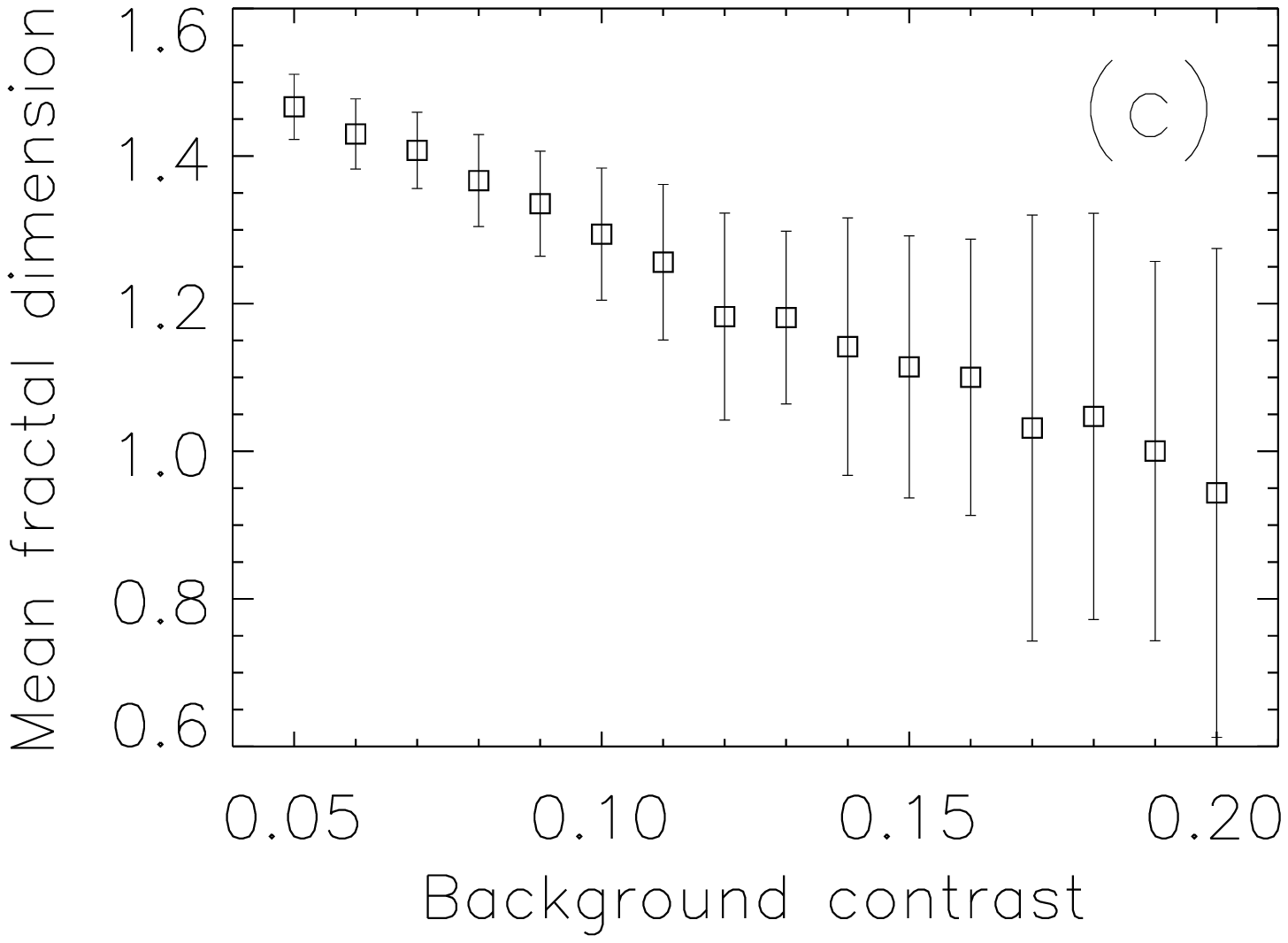}}
\caption{Fractal analysis of the energy release process in EBs 
(from Georgoulis {\it et al.,} 2002):
(a) The calculation of the box-counting fractal dimension $\mathcal{D}$ 
for a given off-band $H\alpha$ image, shown in the inset. 
(b) The calculated fractal dimensions for all $H\alpha$ images and for a 
given contrast threshold $I_{c_0}$. The average fractal dimension for that 
threshold is shown by the dashed line. (c) The dependence of the average 
fractal dimension with the contrast threshold.}  
\label{f2}
\efig
An example of the above analysis in EBs shows that (i) the spatial distribution 
of the released energy is clearly fractal (Figure \ref{f2}a), 
(ii) the fractal dimension varies 
around a well-defined average value for a given contrast threshold used to 
identify EBs (Figure \ref{f2}b), 
and (iii) this average fractal dimension depends 
monotonically on the value of the threshold (Figure \ref{f2}c). 
Dust fractal structures are obtained only for the most stringent threshold used,  
but this threshold is probably too restrictive, so despite 
the magnetic fragmentation intrinsic to 
the EB triggering process, EBs do tend to organize themselves 
in clusters. The physical interpretation of this result lies in the triggering 
locations of EBs and is explained in detail by Georgoulis {\it et al.,} (2002). 
Despite the unambiguous fractality 
of EBs, however, we notice that a quantitative analysis of their fractal dimension 
is essentially threshold-dependent. This is the case for 
other types of sub-flaring activity, as well, and presents a problem in 
the analysis unless one 
has a physically sound reason behind the choice of a given threshold. 
\subsection{Solar Magnetic Fields}
The majority of solar image processing has been applied to solar magnetic fields. 
The observed large-scale, persisting, photospheric magnetic flux accumulations 
that comprise solar ARs allow a detailed study in the course of time. Numerous 
works have shown that the active-region magnetic fields are self-similar 
structures, with area PDFs obeying well-defined power laws and with a fractal 
dimension ranging between $\sim 1.2$ and $\sim 1.7$ (see, e.g., Harvey 1993; 
Harvey and Zwaan 1993; Meunier 1999; 2004; Jan{\ss}en, V\"{o}gler, and 
Kneer 2003 and references therein). The value of the fractal dimension depends on  
whether the structures themselves 
or their boundaries are box-counted. Fractal dimensions in solar magnetic fields 
are typically calculated using the box-counting technique discussed in Section 
2.1. Nevertheless, the analysis has been pursued 
even further, to the concept of multifractality: it is well-known that an AR, 
for example, includes multiple types of structures such as different classes of 
sunspots, plages, emerging flux sub-regions, etc. The physics behind the 
formation and evolution of each of these structures is not believed to be 
identical, so the impact and the final outcome of turbulence in each of these 
configurations should not be the same. For each type of structure being a 
different fractal, an AR is then comprised by an ensemble of fractals, so it 
is a multifractal structure. The fractal dimension of a multifractal set is the 
maximum fractal dimension of the fractal subset that comprise the multifractal 
(Mandelbrot 1983). Numerous studies have revealed the multifractality of 
solar ARs, but also of the Quiet-Sun magnetic fields (e.g., Lawrence, Ruzmaikin, 
and Cadavid 1993; Cadavid {\it et al.,} 1994; Lawrence, Cadavid, and Ruzmaikin 1996; 
Abramenko {\it et al.,} 2001; Abramenko {\it et al.,} 2002). 

Various constructions of multi-scaling, or multifractal, spectra, can be used to 
quantify the multifractal character of solar magnetic fields. In the following, 
we briefly summarize some of these techniques: 
\subsubsection{Generalized correlation dimensions}
The spectrum of generalized correlation dimensions can be derived for both 
timeseries and images (Vlahos {\it et al.,} 1995; Georgoulis, Kluiving, and Vlahos 
1995). One covers the observed magnetic flux image of size $L$ with a uniform 
rectangular grid consisting of elementary boxes with size $\lambda$ and 
dimensionless size $\varepsilon = (\lambda /L)$, as discussed 
in Section 2.1. One then finds the normalized magnetic flux 
$\tilde{\Phi}_i = (\Phi _i / \Phi_{tot})$; $i \equiv 1,...,(N/\lambda)$, for 
each of the boxes $i$, where $\Phi_{tot}$ is the total magnetic flux present 
in the field of view. Since $\sum _i \tilde{\Phi}_i =1$, each normalized flux 
$\tilde{\Phi}_i$ can be called a probability. In case of a multifractal AR, 
the weighted sum of probabilities $\sum _i \tilde{\Phi}_i^q(\varepsilon)$, 
raised to the power $q$ scales with $q$ as follows: 
\beq
\sum _i \tilde{\Phi}_i^q(\varepsilon) \propto \varepsilon ^{(q-1)D(q)}\;\;,
\label{Dq1}
\eeq
where $D(q)$ corresponds to the generalized Renyi dimensions. The ``selector'' $q$ 
is generally a real number, although negative values lack a physical 
interpretation. The function $D(q)$ is typically 
a decreasing function of $q$ for $q>0$. 
For $q=0$ one obtains the fractal dimension $\mathcal{D}=D(0)$ of the 
multifractal, while $D(n)$ for $q=n$ corresponds to the correlation dimension 
between $n$ neighboring flux accumulations, bounded within a box of size 
$\lambda$. If $D(n) <2$, where $2$ is the Euclidean dimension of the image, then 
there is correlation (that is, interaction or clustering) between the neighboring 
flux accumulations. The stronger the decrease of $D(q)$ for increasing $q$, 
the stronger the 
multifractal character of the magnetic flux accumulation. This is because $D(q)$ 
relates to the $q-$tuplet correlation integral $C_q(\varepsilon)$ as 
follows (Hentschel and Procaccia 1983):
\beq
D(q)={{log C_q(\varepsilon)} \over log \varepsilon}\;\;.
\label{Dq2}
\eeq
\subsubsection{Structure functions}
The structure function spectrum (Frisch 1995) has been applied to 
solar magnetic fields by Abramenko {\it et al.,} (2002; 2003) 
in order to quantify the degree of intermittency in solar ARs. Instead of 
box-counting, one now introduces 
a characteristic displacement $\mbf{r}$, called the 
separation vector, and defines a structure function 
\beq
S_q(r) = \langle \mid \Phi (\mbf{x}+\mbf{r}) - \Phi (\mbf{x}) \mid ^q \rangle\;\;,
\label{sf1}
\eeq
where the selector $q$ is also a real, preferably positive, number. 
The notation $\langle P \rangle$ corresponds to a spatial average of the 
quantity $P$ over the field of view. In case of an 
intermittent, multifractal, magnetic flux concentration, $S_q(r)$ exhibits 
a power-law dependence on $r$ within a certain range of displacements, namely,  
\beq
S_q(r) \propto r ^{\zeta (q)}\;\;,
\label{sf2}
\eeq
where $\zeta (q)$ is the exponent of the structure function. In the absence of 
intermittency there is a linear relation $\zeta (q)=(q/3)$ between $\zeta (q)$ 
and $q$, but in actual, intermittent, solar magnetic fields the relation 
between $\zeta (q)$ and $q$ is either nonlinear or it departs significantly 
from the above linear slope. The power-law regime of $S_q(r)$ in which 
$\zeta (q)$ is measured extends between, say, $r=r_1$ and $r=r_2$, $r_1 < r_2$, 
and can be interpreted as the inertial turbulent range where 
free magnetic energy fragments ideally and becomes distributed in 
successively smaller structures. The maximum limit $r_2$ probably refers to 
the maximum size of a magnetic structure that can participate in the 
inertial-range fragmentation, while the minimum limit 
$r_1$ refers to the breakdown of the inertial range and the onset of dissipation 
and subsequent release of free magnetic energy. 

\bfig[t]
\centerline{\includegraphics[width=12.cm,height=10.cm,angle=0]{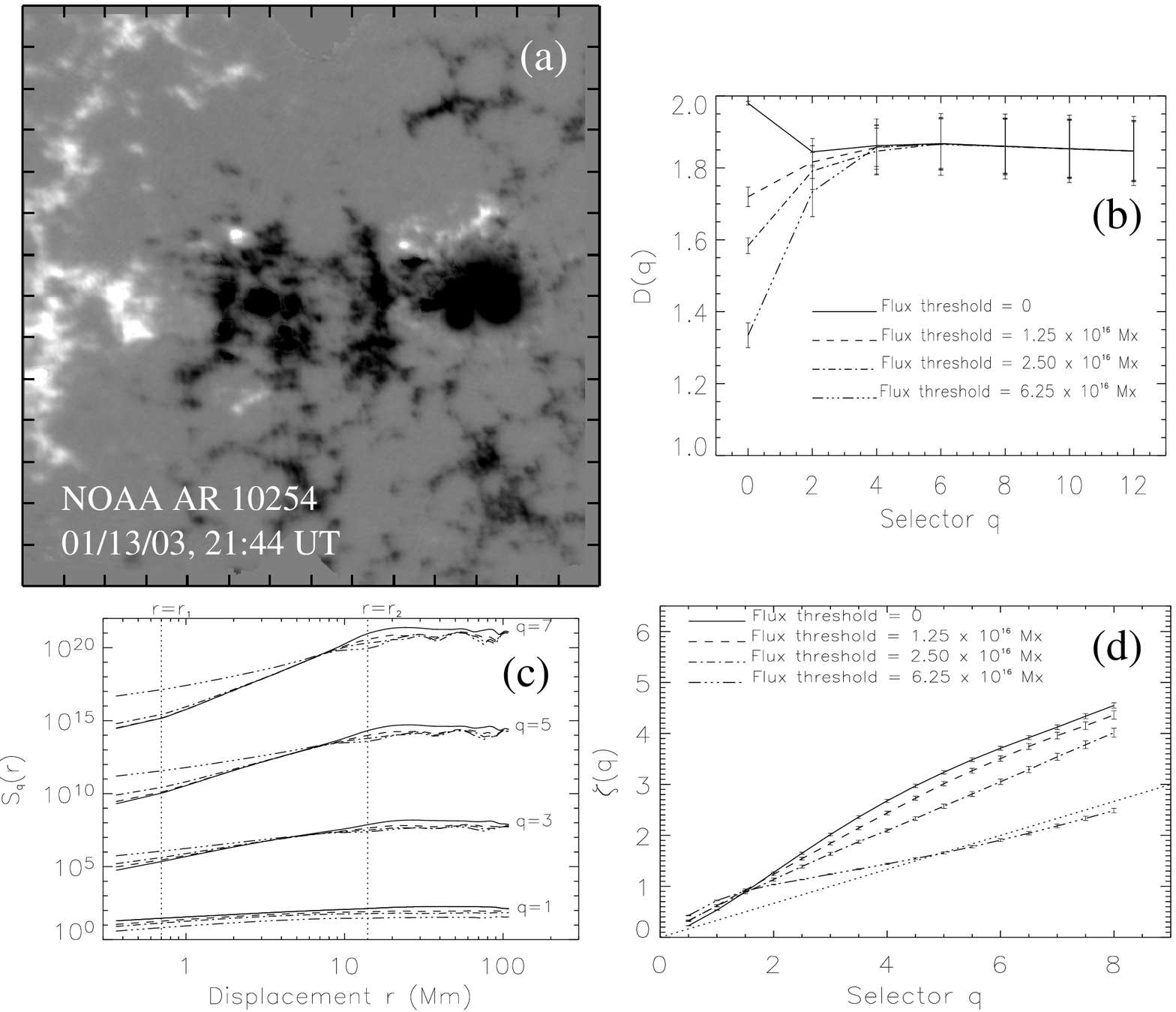}}
\caption{Multi-scaling spectra of NOAA AR 10254 observed by the IVM: 
(a) The measured line-of-sight magnetic field in the AR. The pixel 
size of the IVM image 
is $0.55''$, or $\sim 0.4\;Mm$. Tic mark separation is $20''$. 
Magnetic field values have been saturated at $700\;G$. North is up; west 
is to the right. (b) The spectrum $D(q)$ of the generalized correlation 
dimensions for various selectors $q$. Various magnetic flux thresholds have 
been used in the calculation. (c) The structure function spectrum $S_q(r)$ 
for various displacements $r$ and selectors $q$. The same flux thresholds with 
b have been used. The two parallel dotted lines indicate the common turbulent 
inertial range for all $q$. (d) The structure function exponents $\zeta (q)$ 
for various $q$ and flux thresholds. The dotted straight line corresponds to 
the non-intermittent turbulent case $\zeta (q)=(q/3)$.} 
\label{f3}
\efig
An example of calculation of multi-scaling spectra in solar ARs is given in 
Figure \ref{f3}. The subject is NOAA AR 10254 observed by the Imaging Vector 
Magnetograph (IVM; Mickey {\it et al.,} 1996) of the University of Hawaii on 01/13/03. 
The entire magnetic field vector has been measured by the IVM, so 
the analysis has been applied to the total magnetic flux in the AR. Figure 
\ref{f3}a shows the measured line-of-sight magnetic field. In Figure \ref{f3}b 
we show the spectrum $D(q)$ of the generalized correlation dimensions for various 
$q$ and for various magnetic flux thresholds in the image. These thresholds are 
$(0,\;1.25 \times 10^{16},\;2.5 \times 10^{16},\;6.25 \times 10^{16})\;Mx$ and 
correspond to total magnetic field thresholds $(0,100,200,500)\;G$ per pixel, 
respectively. We notice that $D(q)$ is practically 
independent from the flux threshold for $q \ge 2$. 
Only the fractal dimension $D(0)$ is 
sensitively threshold-dependent, as also found in Section 2.1 (Figure \ref{f2}c) 
for sub-flaring 
activity. Therefore, the multifractal $D(q)$ spectrum is a more robust way 
of studying solar ARs than simply considering the fractal dimension. For 
$q \ge 4$ we notice that $D(q)$ decreases only slightly, if at all, 
indicating a weak multifractal 
character in the AR. In Figure \ref{f3}c we show the structure functions 
$S_q(r)$ calculated for various $q$ and for the flux thresholds defined in 
Figure \ref{f3}b. As with the $D(q)$ spectrum, the structure function $S_q(r)$ 
is fairly independent from the flux threshold used, except for the most stringent 
threshold of $6.25 \times 10^{16}\;Mx$. Moreover, we 
notice that $S_q(r)$ becomes steeper for increasing $q$ and that the inertial 
range $r_1 \le r \le r_2$ is easily discernible for any given $q$. The results 
of Figure \ref{f3}c are in qualitative agreement with the results of Abramenko 
{\it et al.,} (2002). In terms of a quantitative comparison, however, our inertial range 
is shifted to lower displacements with the lower limit $r_1$ extending practically 
down to the pixel size for $q \le 3$. This suggests that magnetograms with better 
spatial resolution will reveal even smaller inertial-scale displacements. 
In Figure \ref{f3}d we show the structure function exponents $\zeta (q)$ for 
various $q$ and using 
the four different flux thresholds. The dotted line indicates 
the expected non-intermittent turbulent spectrum $\zeta (q)=(q/3)$. Only for 
the most stringent flux threshold of $6.25 \times 10^{16}\;Mx$ does the 
$\zeta (q)$ curve resemble the non-intermittent curve. This suggests that most 
of the intermittency in the AR resides in relatively weak 
magnetic fields $B < 500\;G$. We also notice that $\zeta (q)$ increases nearly 
monotonically with increasing $q$, without reaching an apparent saturation value. 
According to Abramenko {\it et al.,} (2003), this reflects a non-flaring period for the 
AR. We will return to this issue in Section 3. 

Other examples of multi-scaling techniques embedded in solar image processing 
include the construction of the wavelet power spectrum. Wavelets have been used 
for diverse purposes but most notably to identify oscillatory wave modes in the 
solar chromosphere (e.g., McIntosh and Smillie 2004; Tziotziou, Tsiropoula, and 
Mein 2004); transition region 
(e.g. De Pontieu, Erd\'{e}lyi, and de Wijn 2003), and the 
corona (e.g. De Moortel and Hood 2000). Wavelets have also been applied to solar 
magnetic fields although in a rather global manner, aiming to explain large-scale, 
universal, solar periodicities, such as the sunspot index and the nature of the 
11-year solar cycle (Polygiannakis, Preka-Papadema, and Moussas 2003). These tasks 
lie beyond the scope of the present study and therefore wavelets will not be 
discussed further. 
\section{Tactical to Practical: Solar Image Processing with Space Weather 
Applications}
We will now examine whether the techniques discussed in Section 2 can be 
useful for flare prediction. In the Introduction we mentioned that flares 
exhibit intermittency both in space and in time. Intermittency 
in a turbulent system prompts one to search for and identify 
{\it critical thresholds} of those few significant degrees of freedom 
that regulate the evolution of the self-organized system. Since solar ARs are 
invariably multifractal magnetic configurations with a variable degree of 
multifractality, it would be interesting to 
investigate critical thresholds in fractal and multifractal diagnostics that 
may distinguish flaring from non-flaring ARs. Such studies require large samples 
of ARs to ensure sufficient statistics, but we will attempt to identify promising 
avenues of further research even with a limited sample of ARs. 
\subsection{Fractal Diagnostics}
\bfig[t]
\centerline{\includegraphics[width=12.cm,height=6.cm,angle=0]{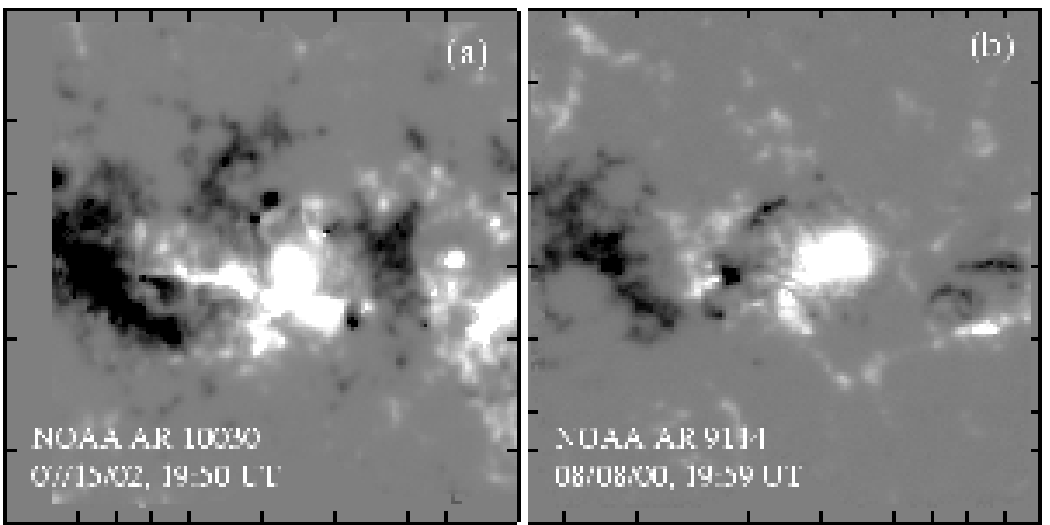}}
\centerline{\includegraphics[width=6.cm,height=4.5cm,angle=0]{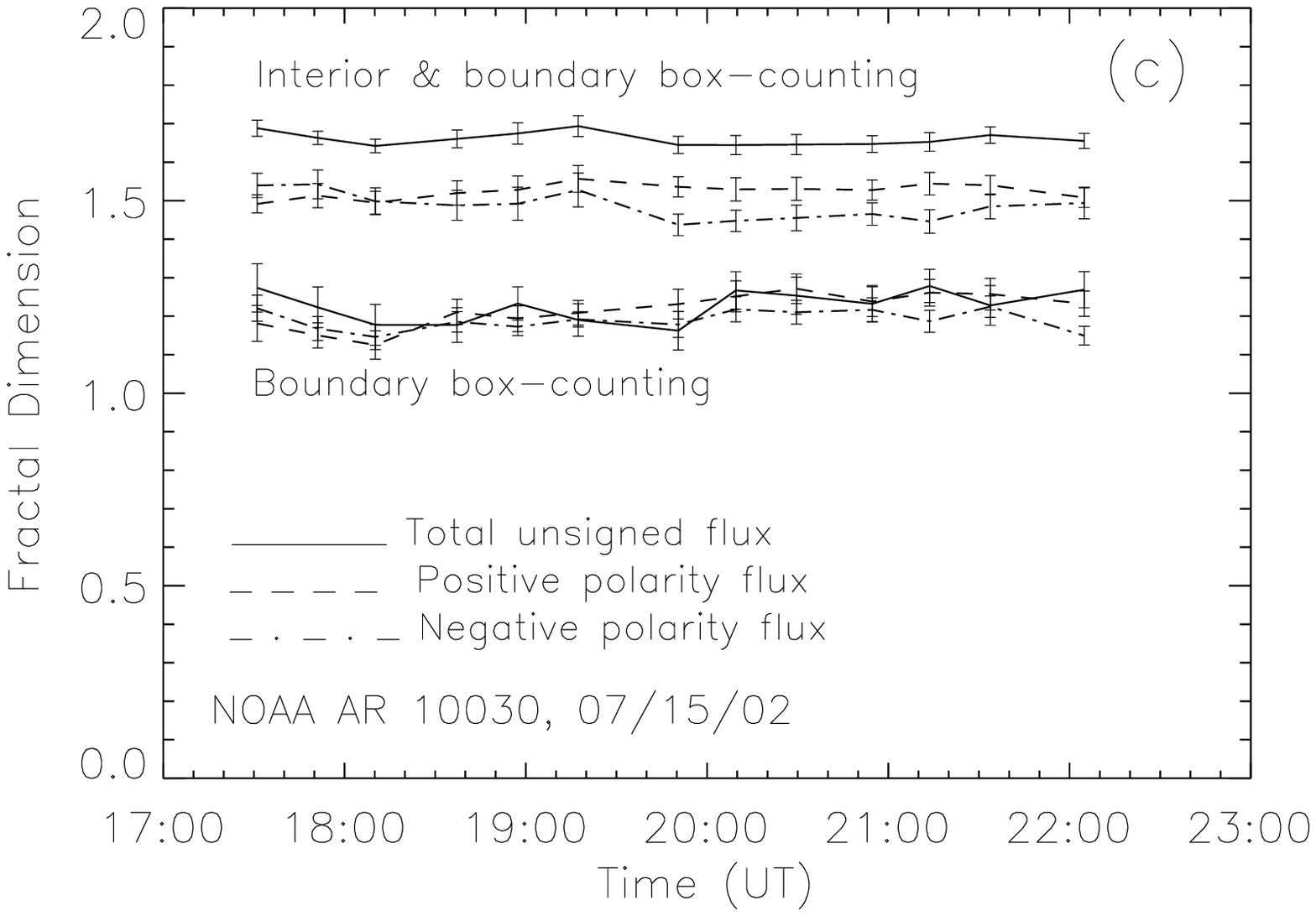}
            \includegraphics[width=6.cm,height=4.5cm,angle=0]{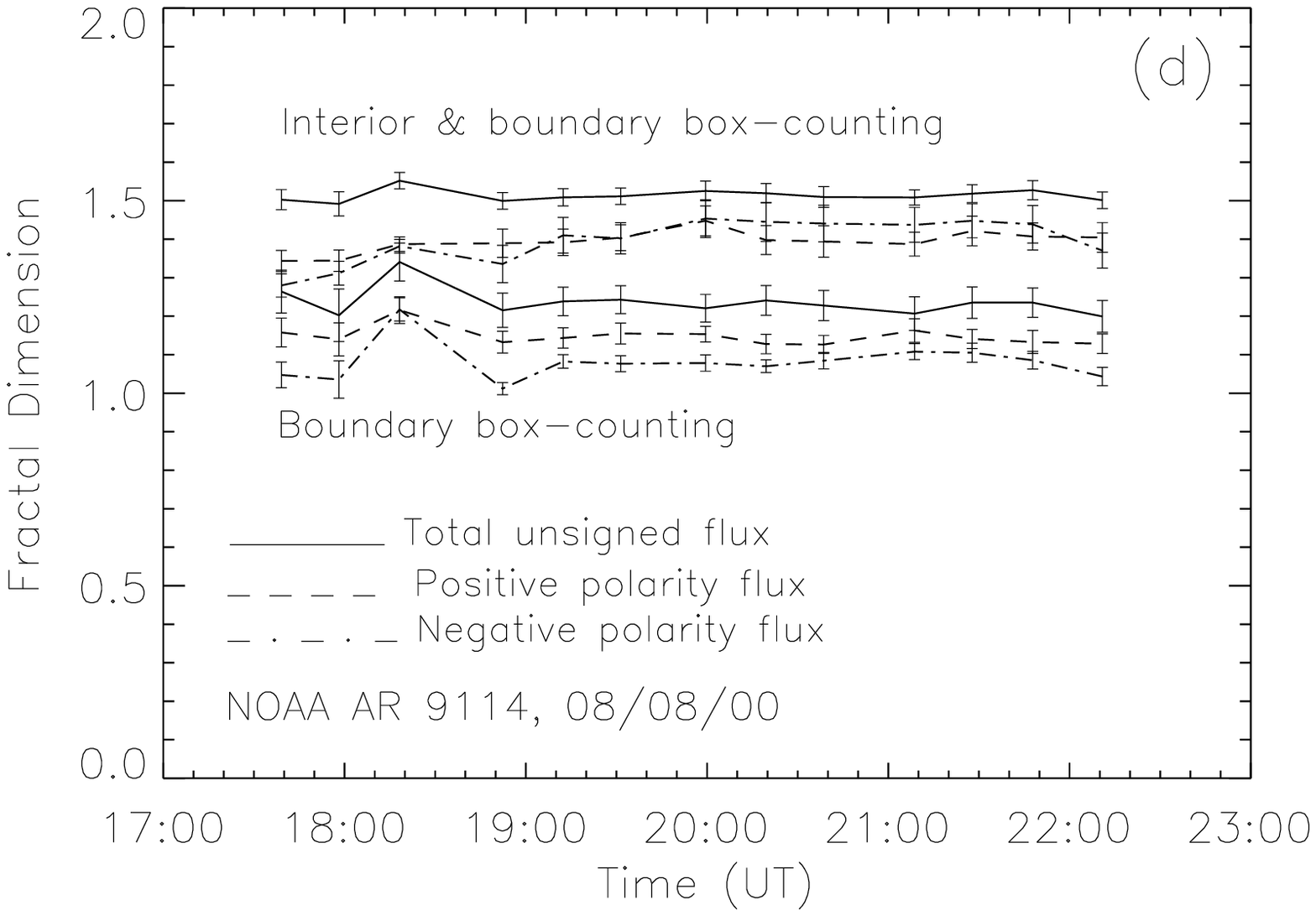}}
\caption{A fractal dimension comparison between a flaring and a quiescent solar AR. 
Both ARs have been observed by the IVM. The fractal analysis in box ARs 
has been performed using a flux threshold of $1.25 \times 10^{16}\;Mx$. 
(a) NOAA AR 10030, which gave a X3 and a M1.8 flares during 
the IVM observations. (b) NOAA AR 9114, with no associated flaring activity. 
Tic mark separation in (a) and (b) is $20''$. North is up; west is to the 
right in both images. (c) Various measures of the fractal dimension for 
NOAA AR 10030. (d) Various measures of the fractal dimension for NOAA AR 9114.}
\label{f4}
\efig
A flaring and a quiescent ARs observed by the IVM are given in Figures \ref{f4}a 
and \ref{f4}b, respectively. NOAA AR 10030 gave two major flares, 
namely a X3 and a M1.8, 
during the IVM observing interval. 
NOAA AR 9114 gave no flares, not even C-class events, 
on the particular day of the observations. The two ARs are distinctly different 
in terms of magnetic complexity and flow patterns, and show 
a factor of two difference in total magnetic flux ($\sim 8 \times 10^{22}\;Mx$ and 
$\sim 4 \times 10^{22}\;Mx$ for ARs 10030 and 9114, respectively). We have 
measured various fractal dimensions for the two ARs over the course of the IVM 
observations. Since the values of the fractal dimension are threshold-dependent 
(Figures \ref{f2}c, \ref{f3}b), we have chosen the same magnetic flux threshold 
of $1.25 \times 10^{16}\;Mx$ ($100\;G$ per pixel) for both ARs and at any given 
time. We have measured the fractal dimension of the magnetic patches and their 
boundaries, both using the unsigned total magnetic flux and discriminating between 
positive and negative magnetic polarity. Our results are given in 
Figures \ref{f4}c and \ref{f4}d for ARs 10030 and 9114, respectively. By 
inspecting and comparing the two plots we reach the following conclusions: 
\ben
\item[(1)] The values of the various fractal dimensions fluctuate slightly in 
the course of time, around a well-defined average value. This is the 
case for both ARs. Moreover, there are no measurable changes in any fractal 
dimension that reflect the triggering of the two flares 
in AR 10030. The onset times of 
the flares were 20:02 UT and 21:30 UT for the X3 and the M1.8 flare, respectively. 
\item[(2)] The fractal dimension measured when both the interior and the 
boundaries of the patterns are box-counted is higher than that measured from only 
the boundaries of the patterns. Discriminating between the two polarities does 
not alter this conclusion. Both ARs 
give boundary fractal dimensions varying between $1.1$ and $1.3$. AR 10030, 
however, gives larger fractal dimensions when boundaries and interiors are 
box-counted.. In AR 10030 the values of the fractal 
dimension range between $1.5$ and $1.7$, while in AR 9114 they vary 
between $1.3$ and $1.5$. This feature is {\it not} related 
to the flare productivity of AR 10030: this region is spatially more 
extended compared to AR 9114, so it occupies a larger fraction of the fixed IVM 
field of view. Since the interiors of the flux patterns are box-counted, 
there are more filled elementary boxes for AR 10030 than for AR 9114 
which increases the fractal dimension. In view of this effect, which might lead 
to misinterpretation of the results, it appears safer to use the boundary fractal 
dimensions. 
\een

I general, we find no distinguishing feature in the measurement of the fractal 
dimension that can hint about the dramatically different activity levels in 
NOAA ARs 10030 and 9114. The values of the various fractal dimensions are similar 
for both ARs, while in cases where there is a difference it is not related to  
the flare productivity in the AR. Moreover, no noticeable changes in the fractal 
dimension occur before and after the flares. In addition, given the fact that the 
values of the fractal dimension depend sensitively on employed flux threshold, we 
conclude that the fractal dimension is not a useful means of distinguishing 
flaring from quiescent ARs, although it quantifies the degree of self-similarity 
in solar ARs. We cannot rule out the possibility that statistical patterns might 
be obtained when the fractal dimension is calculated for a large number of 
flaring and quiescent ARs, but this example here indicates that there might be 
little hope for that. The fractal dimension can be useful for other 
types of studies, nevertheless, such as those dealing with 
the different phases of the solar cycle. 
Meunier (2004) reports a significant variation of the fractal dimension between 
solar minimum and solar maximum. However, the objectives of space weather 
forecasting are different and essentially focused on short- and mid-term 
predictions of the eruptive potential in solar ARs, rather than on long-term 
predictions of the order of the solar cycle.  
\subsection{Multifractal Diagnostics}
Let us now investigate whether multifractal spectra can help one distinguish 
between flaring and quiescent ARs. For this task, we have chosen a sample of 
six ARs observed by the IVM. Three of these ARs, namely NOAA ARs 10030, 8210, 
and 9165, gave at least M-class flares on the day of the IVM observations and 
they are considered ``flaring'' ARs. The remaining three ARs, namely NOAA ARs 
9114, 8592, and 10254, are not associated with M-class or larger flares 
on the day of the 
observations and they are considered ``quiescent'' ARs. Although we have shown 
in Section 2.2 that the multifractal spectra do not depend sensitively on the 
employed flux threshold, for comparison purposes we 
use the same threshold for all six ARs. 
This flux threshold is $1.25 \times 10^{16}\;Mx$. 

\bfig[t]
\centerline{\includegraphics[width=10.cm,height=7.cm,angle=0]{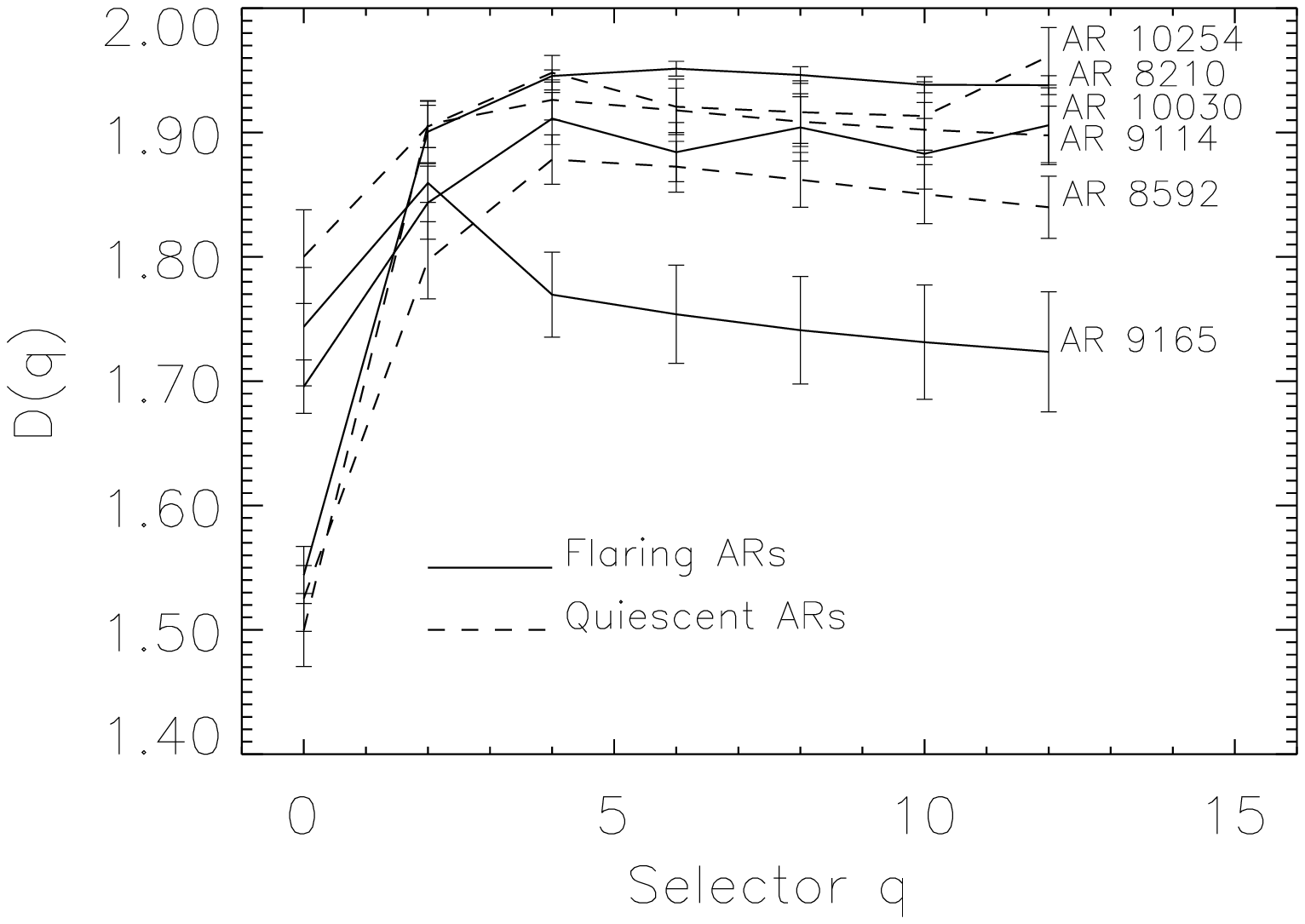}}
\caption{Generalized correlation dimensions $D(q)$ for various selectors $q$ 
and for a sample of flaring (solid curves) and non-flaring (dashed curves) ARs.} 
\label{f5}
\efig
We first calculate the generalized correlation dimensions $D(q)$ for all ARs and 
for various selectors $q$. The results are shown in Figure \ref{f5}. It is 
evident that there is no distinguishing feature between the flaring and the 
non-flaring ARs, since the shape of the $D(q)$ spectra and the $D(q)$-values 
are similar, within error bars, for both types of ARs. The exception from this 
conjecture is AR 9165 with significantly lower $D(q)$-values compared to the 
other ARs. This implies that AR 9165 exhibits the strongest multifractal 
character and it is also a flaring AR. Nonetheless, we cannot 
definitively link the degree 
of multifractality with the flare productivity, since the other two flaring ARs 
do not show this feature. AR 9165 was not even the most 
flare-productive AR. On the day of the observations AR 9165 gave a M2 flare, 
while AR 10030 gave a X3 and a M1.8 flare. In conclusion, the construction of the 
generalized correlation dimensions $D(q)$ is not particularly useful in 
identifying potentially flaring ARs. 

We then calculate the structure functions $S_q(r)$ and their inertial-range 
scaling exponents $\zeta (q)$ for various selectors $q$ and displacements $r$. 
This is a very interesting test given that there are reports in the literature 
arguing that $S_q(r)$ and $\zeta (q)$ can provide clues for the flare productivity 
of solar ARs. In particular, Abramenko {\it et al.,} (2002; 2003) report that (i) the 
$S_q(r)$ spectra are flatter for flare-productive ARs compared to quiescent ARs, 
and (ii) the shape of the $\zeta (q)$ spectrum deviates significantly from the 
linear non-intermittent case, $\zeta (q)=(q/3)$, for flaring ARs. The flattening 
of the $S_q(r)$ spectra and the nonlinear $\zeta (q)$ curve suggest an increase 
of intermittency and they are reported to occur in ARs prior to a solar 
flare. 

\bfig[t]
\centerline{\includegraphics[width=6.cm,height=4.5cm,angle=0]{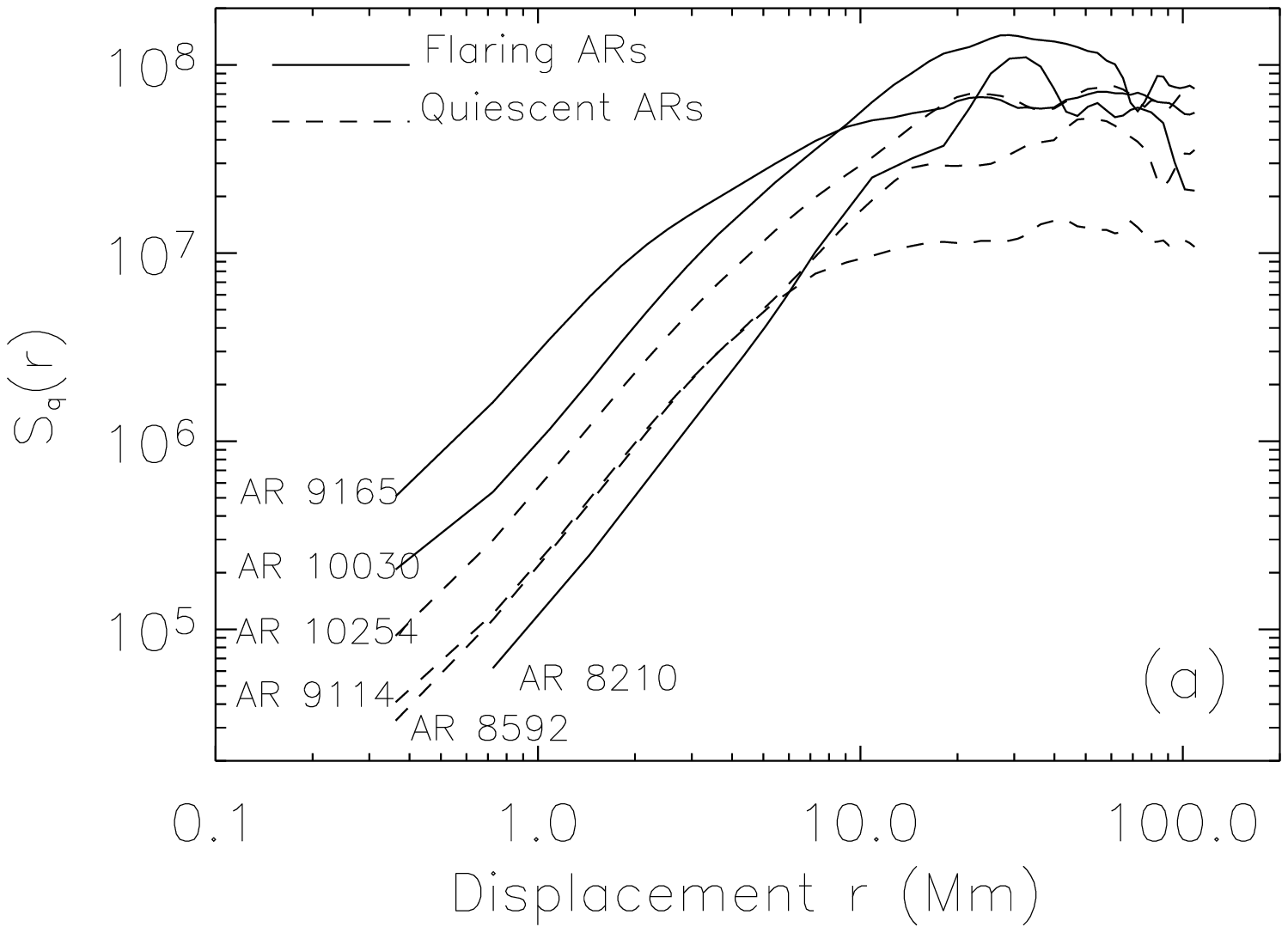}
            \includegraphics[width=6.cm,height=4.5cm,angle=0]{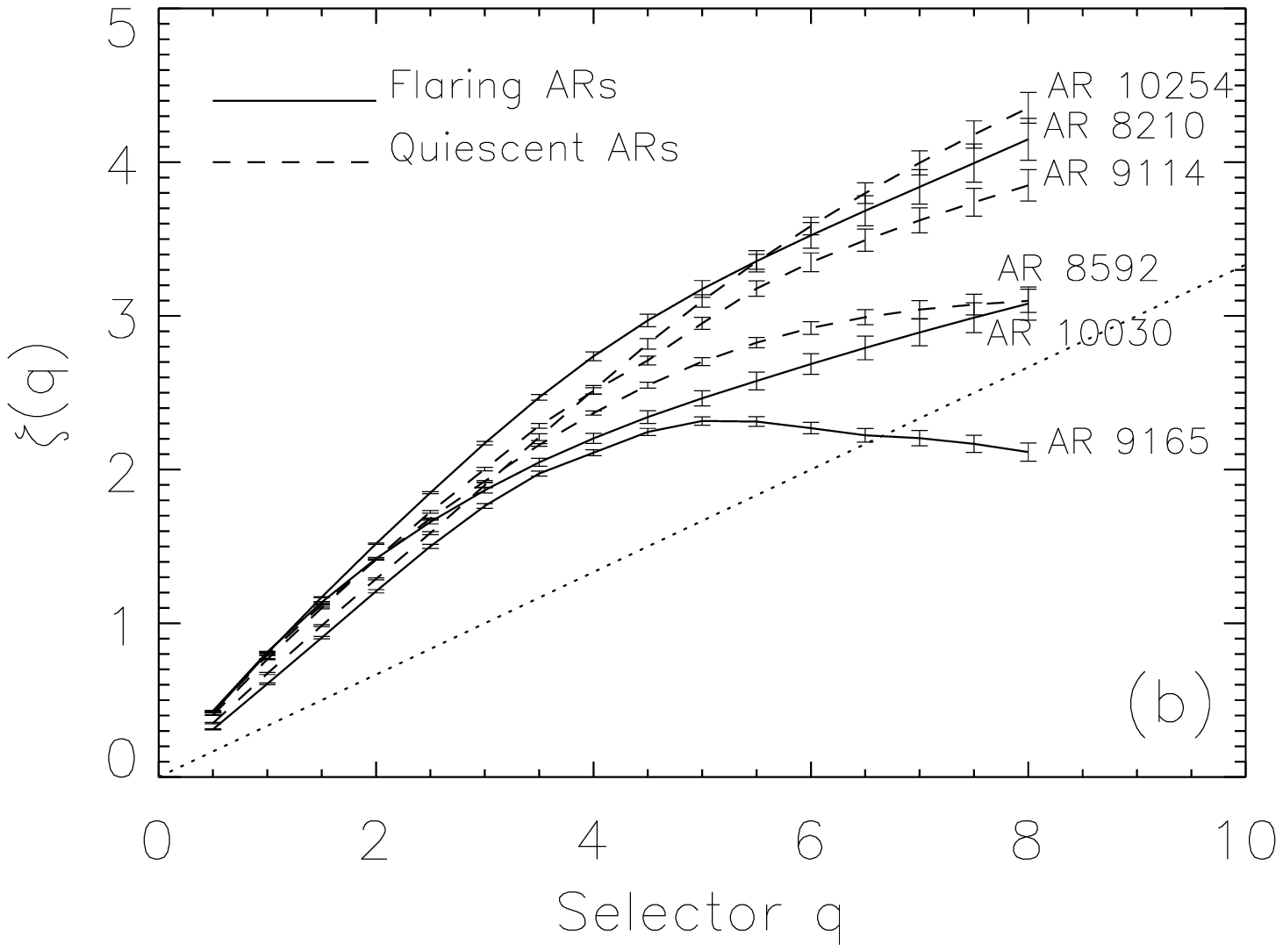}}
\caption{Structure function calculations for a sample of flaring (solid curves) 
and non-flaring (dashed curves) ARs. (a) The structure functions $S_3(r)$ for 
$q=3$. The power-law regimes indicate the turbulent inertial range of the magnetic 
fragmentation process in the ARs. (b) The inertial-range scaling exponents 
$\zeta (q)$ of the structure functions $S_q(r)$ for various selectors $q$. 
The dotted straight line corresponds to the expected spectrum, $\zeta (q)=(q/3)$, 
in case of a non-intermittent turbulence.}    
\label{f6}
\efig
The $S_3(r)$ values ($q=3$) for all chosen ARs are given in Figure \ref{f6}a. 
Other values of the selector $q$ give similar results. 
We notice that, with the exception of AR 8210, the flaring ARs tend to have 
larger inertial-range values of $S_3(r)$ for a given $r$. It is not clear 
why AR 8210 does not follow this trend and we certainly cannot rule out the 
possibility of a random occurrence of this result 
given the limited sample of ARs. Nevertheless, it appears worthy to perform 
the same study for a much larger number of ARs aiming to 
investigate whether flaring ARs tend to give statistically larger values of 
$S_q(r)$ for given $q$ and $r$. The shape of the $S_q(r)$ spectrum does not 
appear to provide any other distinguishing feature between flaring and 
non-flaring ARs. In Figure \ref{f6}b, on the other 
hand, we find that two out of three flaring ARs (ARs 10030 and 9165) show smaller 
$\zeta (q)$ exponents for $q \ge 4$ compared to the other ARs, suggesting that 
the $S_q(r)$ are flatter in their inertial range for these two cases. This is 
in agreement with Abramenko {\it et al.,} (2002; 2003). AR 8210 again exhibits an unclear 
behavior with the values of $\zeta (q)$ being comparable to those of the quiescent 
ARs. Moreover, it is also troubling that AR 9165 gives the strongest nonlinear 
response for $\zeta (q)$ although it is not the most flare-productive AR. In 
qualitative agreement with Abramenko {\it et al.,} (2002; 2003), on the other hand, 
the shape of the $\zeta (q)$ spectrum in flaring ARs deviates more from the 
non-intermittent linear 
case (dotted line) than in non-flaring ARs. This conclusion 
does not include the flaring AR 8210. 

Summarizing, we generally confirm the results of Abramenko {\it et al.,} (2002; 2003) 
but not without exceptions. Clearly, the proposed multi-scaling criteria cannot 
discriminate unambiguously between flaring and quiescent ARs and the diagnostics 
are not more profound for the most flare-productive ARs. The uncovered trends, 
however, make it worthy to study the $S_q(r)$ and $\zeta (q)$ spectra for a large 
number of subject ARs aiming to reveal statistical, probabilistic, patterns 
regarding the flare productivity in ARs. Our analysis suggests that emphasis 
should be placed on the following aspects:
\ben
\item[(1)] The inertial-range values of the $S_q(r)$ spectra for a 
given $q$ (flaring ARs may tend to give larger $S_q(r)$), 
\item[(2)] The values of 
$\zeta (q)$ and the shape of the $\zeta (q)$ spectrum (flaring ARs may tend to 
give more nonlinear $\zeta (q)$ response with smaller values of $\zeta (q)$ for 
larger $q$). 
\een
If such patterns are statistically confirmed, the one might be able 
to define critical thresholds $S_{q;cr}(r)$ and $\zeta _{cr}(q)$ for 
$S_q(r)$ and $\zeta (q)$, respectively. Values $S_q(r) > S_{q;cr}(r)$ for 
given $q$ and $r$, as well as values $\zeta (q) < \zeta _{cr}(q)$ for a given 
$q$, might imply an enhanced likelihood of flaring events in an AR. 

We note in passing that the same sample of ARs has been used in different 
investigations 
of the flaring vs. the quiescent activity. In Georgoulis and LaBonte (2005) 
we provide more clear distinguishing features between flaring and non-flaring ARs 
based on the free magnetic energy and the total magnetic helicity budgets of 
the above sample of ARs. 

A further important test is to examine the multifractal behavior of a flaring AR 
in the course of time and ideally before and after the flaring event(s). As 
discussed in Section 3.1, NOAA AR 10030 is appropriate for this task since it 
produced two flares during the IVM observing interval. The flare onset times were 
20:02 UT and 21:30 UT for a X3 flare and a M1.8 flare, respectively. We selected 
4 snapshots of the AR before the X3 flare (17:31 UT; 18:10 UT; 18:57 UT; and 
19:50 UT) and 3 snapshots after the event (20:29 UT; 21:13 UT; and 22:04 UT). 
We carefully excluded from the analysis magnetograms obtained during the rising 
phase and the peak of the flares to avoid possible contamination of the 
polarization measurements from the white-light 
flare emission (see, e.g., Qiu and Gary 2003). 
Moreover, we used a fixed flux threshold for all magnetograms, equal to 
$1.25 \times 10^{16}\;Mx$. 

We first examined the temporal behavior of the generalized correlation 
dimensions $D(q)$. This test did not provide useful results. 
In fact the only result was a decrease of the $D(q)$ values 
after the X3 flare at 20:29 UT which, however, was insignificant and mostly kept 
within error bars. This implies that the degree of multifractality in the AR 
increased after the X-class flare. This increase, however, is not 
significant enough to overcome error bars and become useful for flare prediction. 

\bfig[t]
\centerline{\includegraphics[width=6.cm,height=4.5cm,angle=0]{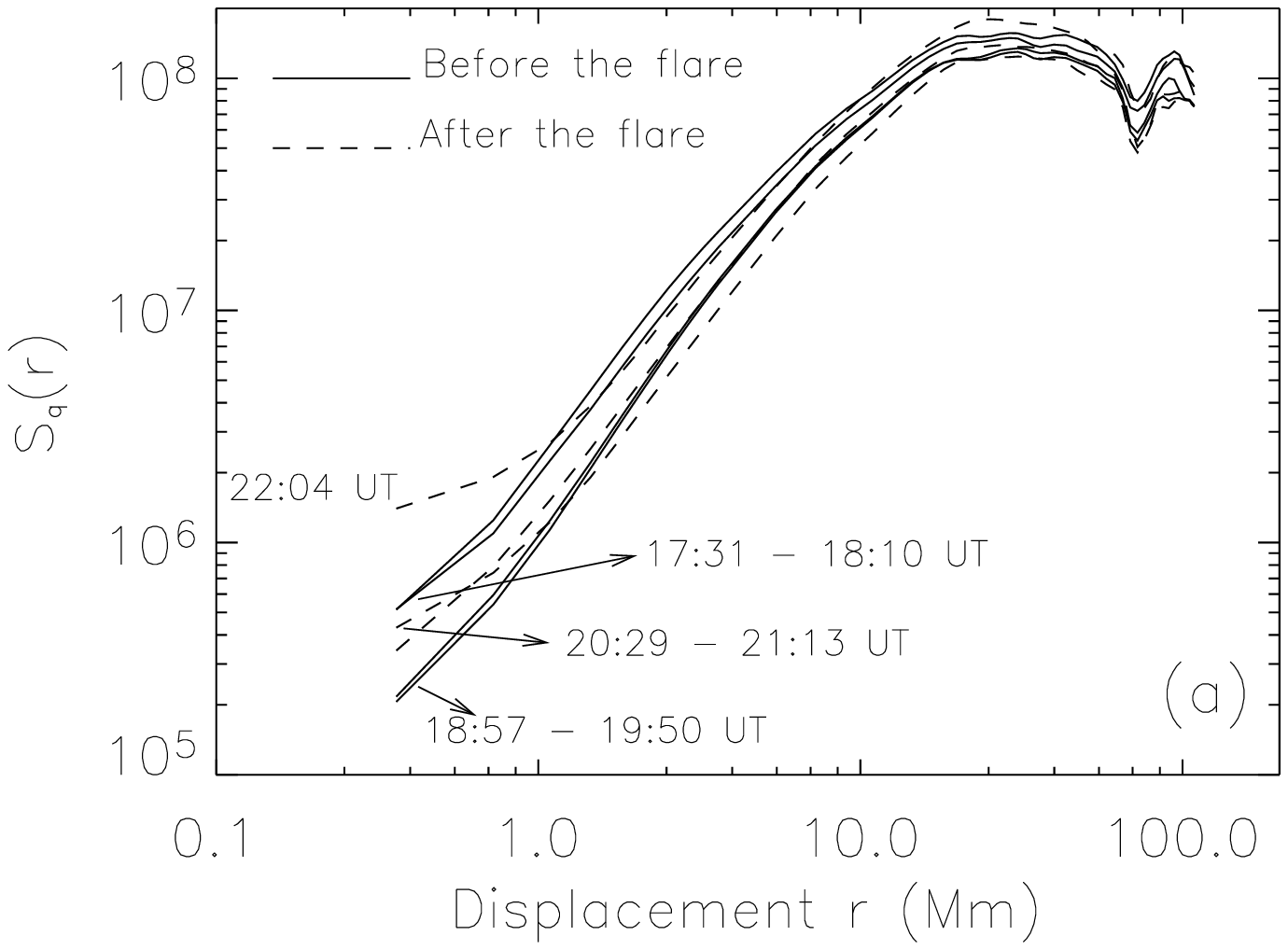}
            \includegraphics[width=6.cm,height=4.5cm,angle=0]{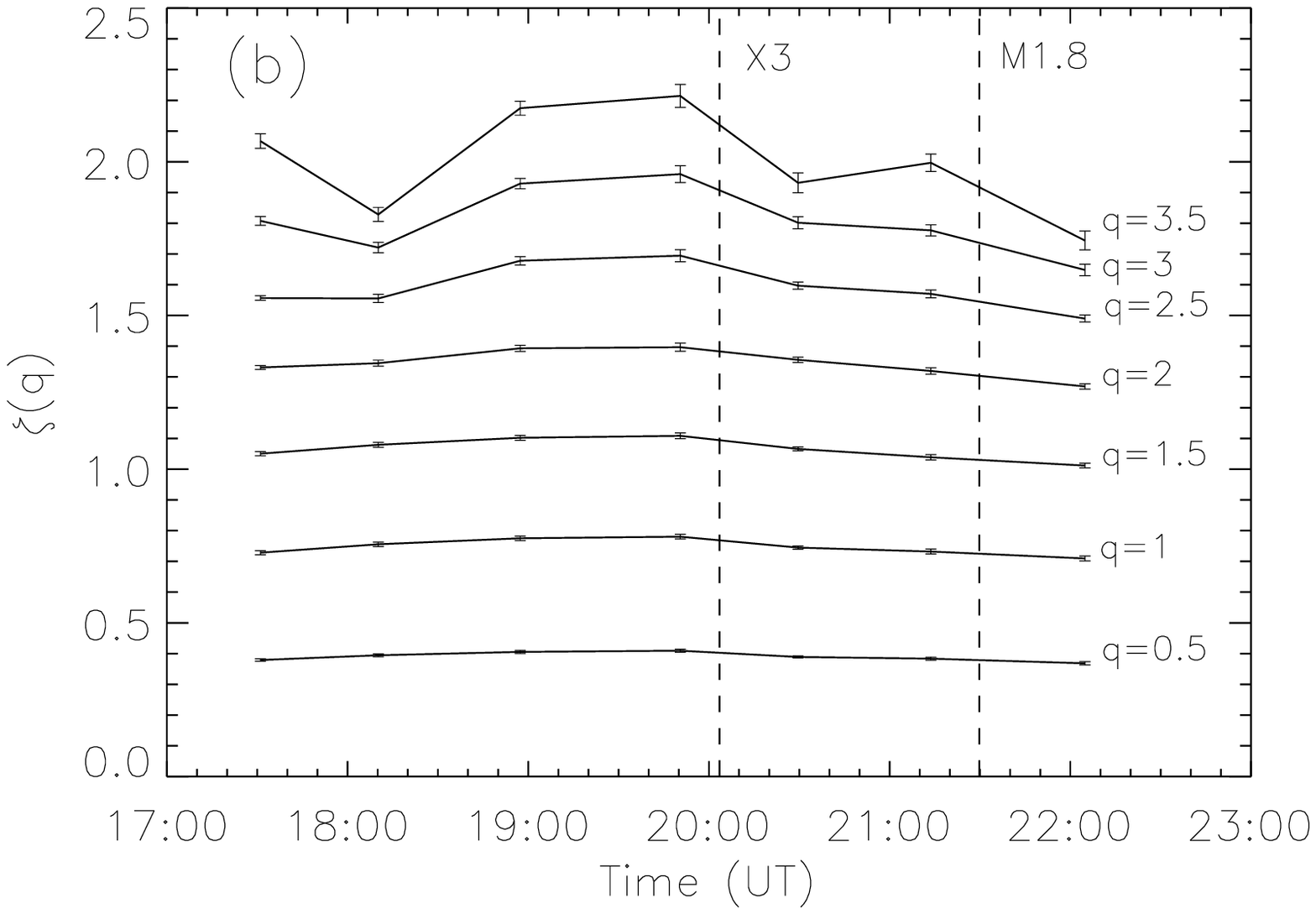}}
\caption{Temporal multifractal analysis of the flaring NOAA AR 10030. 
(a) The structure functions $S_3(r)$, for $q=3$, for various times before 
and after the X3 flare (solid lines and dashed lines, respectively). 
(b) The temporal evolution of the inertial-range scaling exponents $\zeta (q)$ 
for various selectors $q$. The onset times of the two flares triggered in the 
AR are indicated by the two parallel dashed lines.} 
\label{f7}
\efig
We then calculated the structure functions $S_q(r)$ and their inertial-range 
exponents $\zeta (q)$ as functions of time. The values of $S_3(r)$ for $q=3$ 
and for various displacements $r$ are given in Figure \ref{f7}a. Similar 
results are obtained for other choices of $q$. From Figure 
\ref{f7}a we notice that (i) there is no significant difference in the 
inertial-range values of $S_q(r)$ before and after the flares, and (ii) the 
inertial power-law regime {\it flattens} after the events. The effect is 
more discernible in Figure \ref{f7}b and for $q \ge 2$, where the values of 
$\zeta (q)$ decrease after the events. As a result, the response of $\zeta (q)$ 
becomes more nonlinear after the flares, suggesting that the degree of 
intermittency in the ARs has been increased after the events. This stands in 
agreement with the results of Abramenko {\it et al.,} (2002; 2003). The decrease of 
$\zeta (q)$ for a given $q \ge 2$ is beyond error bars and ranges from 
$\sim 4\%$ ($q=2$) to $\sim 20\%$ ($q=3.5$). We have performed the same test 
for the quiescent AR 9114 and we could not detect such changes of $\zeta (q)$ 
in the course of time. The temporal variation of $\zeta (q)$, therefore, 
can be used to identify flaring periods in an AR. To become even more practical 
for flare prediction, however, $\zeta (q)$ should also be able to identify 
{\it pre-flare} periods in solar ARs. From Figure \ref{f7}b we notice an 
increase of $\zeta (q)$ before the X3 flare and for $q \ge 2$. 
This behavior suggests a pre-flare 
decrease of the intermittency in the AR and is discernible after 18:00 UT, 
that is $2\;hr$ before the X3 flare. 

In conclusion, the temporal variation of $\zeta (q)$ in a flaring AR is probably 
the most useful multifractal diagnostic of imminent flaring activity. Further 
tests on large numbers of flaring and non-flaring ARs are required to substantiate 
this. The tale-telling signature for an imminent flare appears to be a significant 
increase in $\zeta (q)$ a few hours before the event followed by an also 
significant decrease of $\zeta (q)$ after the event. The sufficient statistics 
of a comprehensive analysis should be able to establish critical thresholds of 
the pre-flare increase of $\zeta (q)$ beyond which a flare might 
be safely predicted. 

More multifractal indices can be defined as by-products of the multi-scaling 
analysis in solar ARs (see Abramenko {\it et al.,} 2003 for examples) but one should 
focus on the right combination of simplicity and usefulness in the construction 
and practical use of a given index for space weather purposes. 
From this aspect, the inertial-range 
exponents $\zeta (q)$ of the structure functions $S_q(r)$ are probably the most 
appropriate multifractal diagnostics of flares, 
given their simple construction and sound physical justification. 
\subsection{Other Image Processing Techniques with Space Weather Applications}
In this study we focused on image processing techniques directly stemming from 
the theory of nonlinear dynamical systems exhibiting intermittent 
self-organization.  This is not the only image processing applied to solar data. 
Additional solar image processing techniques with an impact in space weather 
forecasting generally refer to two broad categories: (1) pattern recognition, and 
(2) contrast enhancement and difference imaging. 

Pattern recognition techniques apply to active-region recognition and 
sunspot identification / classification on the visible solar disk (e.g. 
Turmon, Pap, and Mukhtar 2002; see also the web pages of the Active Region 
Monitor and the Max Millenium Program of Solar Flare Research), filament 
identification from $H \alpha$ images of the solar chromosphere (Bernasconi, 
Rust, and Hakim 2005; this volume), sigmoid recognition from soft X-ray 
images of the lower solar corona (LaBonte, Rust, and Bernasconi 2004), and 
CME recognition from white-light images of the outer corona 
(Robbrecht and Berghmans 2004 and in this volume). Sunspot 
and active-region identification is a prerequisite for the automatic application 
of the scaling and multi-scaling techniques discussed in the previous Sections 
and hence this task is an integral part of the solar research required for 
space weather forecasting. Equally important is the automatic, real-time, 
identification of CMEs. Filament and sigmoid recognition, on the other hand, 
can help understand 
the pre-launch phases of a CME. Although these techniques are not directly 
related to the theory of intermittent turbulence, there is a critical threshold 
that can place the CME prediction on a quantitative basis. This threshold comes 
from the helical structure of solar magnetic fields and the flux ropes that fuel 
a CME. If the aspect ratio $(L/R)$ of a, say, cylindrical flux rope is larger than 
a critical threshold $(L/R)_{cr}$, then the flux rope becomes kink-unstable and 
may erupt. In the above notation $L$ is the length of the flux rope and $R$ is the 
radius of the cylinder. Rust and Kumar (1996) have shown that 
$(L/R)_{cr} \simeq 5.4$. Measurement of the aspect ratios of 
X-ray sigmoids can be performed automatically and in real time. 
Moreover, the shape of 
filaments and sigmoids can reveal the sense of magnetic helicity (chirality) 
of the pre-eruption structure thus adding to the predictive ability of these 
pattern recognition techniques. The notion of a critical threshold provided by 
the theory of the kink instability fits nicely with the viewpoint 
that CMEs are caused by a loss of equilibrium in the solar atmosphere 
(see Forbes 2000 and references therein). The CME phenomenon is also characterized 
by temporal intermittency and self-similarity with the CME launch lasting only 
a few $hr$ and the PDFs of the CME kinetic energies obeying well-defined 
power laws (Vourlidas 2004; private communication). Therefore, 
it is conceivable that SOC models and fractal/multifractal diagnostic techniques 
may be developed in the future to study the CME initiation process. 

Contrast enhancement and difference imaging apply to the detection of EUV 
transient dimmings that accompany the launch of a CME (e.g. Aschwanden 2005; 
this volume), the detection of Moreton waves, or ``EIT waves'', following 
a flare and/or a CME (Thompson {\it et al.,} 1999), and the sharpening of the white-light 
CME images observed by SoHO/LASCO (Stenborg and Cobelli 2003; Stenborg 2005; this 
volume). Transient dimmings accompanying the CME initiation are long known and 
frequently observed (e.g. Rust 1983). They are thought to correspond to the 
footpoints of the CME and the subsequent interplanetary flux rope (e.g. Kahler 
and Hudson 2001 and references therein). Therefore, their detection is of 
particular space weather importance. The physical nature of the EIT waves remains 
elusive (Zhukov and Auch\'{e}re 2004) but their study, via the study of 
differences between consecutive EUV images, contributes to the 
construction of a consistent and comprehensive physical picture of the CME 
phenomenon. Regarding the sharpening of the CME images using, for example, 
wavelet techniques, we suggest that a semi-quantitative analysis of the magnetic 
helicity of CMEs can be performed, although it is questionable whether this 
analysis can be done automatically. In particular, one can arguably measure the 
amount of twist, via the length of the CME structure and the number of turns 
as they are revealed from the sharpened image, and the amount of writhe, thus 
estimating the total magnetic helicity of the structure. This estimation can be 
compared against the estimated helicity of the pre-CME sigmoids and filaments 
in order to understand which magnetic fields contribute to a CME. 
The latter is a completely open 
question, with the debate holding on whether CMEs are local events, i.e., 
they originate from single active-region magnetic fields/filaments, 
or global events, i.e., 
requiring an interaction between the active-region magnetic fields and the global 
solar dipole. Comparable helicities between the 
CME and the source sigmoid, for example, would point to a direct relationship 
between the CME and the X-ray coronal loop, while large discrepancies would 
suggest that CMEs are events with their magnetic structure 
contributed by the large-scale solar magnetic fields besides the source magnetic 
structure. CME image enhancement has already uncovered the rich structure 
of the CMEs' magnetic configuration, but via this plausible research option we 
might be able to gain much more understanding of the CME triggering process.  
\section{Summary and Conclusions}
Intermittent MHD turbulence is a key feature of magnetic fields, their 
evolution, and the associated eruptive activity in the Sun's atmosphere. 
As a result, fractal and multifractal image processing techniques are applicable 
and can be used to quantify the degree of intermittency, self-similarity, 
and multifractality present in the system. While these multi-scaling techniques 
can advance our understanding of solar magnetic fields, it is questionable 
whether they can be used to predict eruptive activity such as solar flares and 
CMEs. This is because intermittency, self-similarity, 
and multifractality are so widespread 
in the solar magnetic fields that it is not clear whether the above techniques 
are appropriate to discriminate between different types 
(i.e. flare-productive and quiescent) magnetic configurations. 

We address this question in the present study. In particular, we ask whether 
fractal and multifractal diagnostics can be of any predictive importance. 
We focus on the prediction of solar flares, rather than on CME prediction, 
because there are several open questions regarding the origin of CMEs that 
do not allow the zero-level understanding required to adjust multi-scaling 
techniques for their study. Prompted by the intermittency of the solar flare 
phenomenon and the overall self-organization of the solar magnetic fields 
we search for possible critical thresholds in the scaling and multi-scaling 
behavior of the nonlinear dynamical system which, if exceeded, can give rise 
to intermittent energy dissipation, namely, to a solar flare. 
To study flare-productivity in solar ARs we measure the 
fractal dimension, the multifractal generalized correlation dimensions, 
the multifractal structure functions, and their inertial-range scaling exponents 
of the photospheric magnetic flux comprising these ARs. 
Our results can be summarized as follows: 
\ben
\item[[1]] The calculation of the fractal dimension is not a particularly 
fruitful way to discriminate flaring from quiescent ARs. The value of the fractal 
dimension is sensitively threshold-dependent and is similar between the two 
different types of ARs. Moreover, care is required to the definition of the 
fractal dimension. When both the boundaries and the interiors of magnetic flux 
patterns are used in the measurement of the fractal dimension, then the results 
become dependent on the ratio between the spatial extent of the AR and the 
finite field of view. This can lead to misinterpretation of the results and in 
differences in the values of the fractal dimension that are not related to 
flaring activity. Therefore, we expect that the use of the fractal dimension for 
flare prediction purposes should be limited. 
\item[[2]] The calculation of the generalized correlation dimensions $D(q)$ 
is also not particularly useful. Flaring and non-flaring ARs tend to show similar 
values of $D(q)$. The gain in the use of the $D(q)$ spectrum is that it reveals 
the degree of multifractality of active-region magnetic fields. However, this 
does not appear to be a sensitive function of the flare productivity in solar 
ARs. Therefore, $D(q)$ spectra are not expected to improve our flare-forecasting 
ability. 
\item[[3]] The calculation of the structure functions $S_q(r)$ and their 
inertial-range exponents $\zeta (q)$ have provided some forecasting clues, 
but not without limitations. In particular, (i) flaring ARs tend to show larger 
inertial-range $S_q(r)$ 
values for given $q$ and $r$, (ii) flaring ARs tend to have flatter 
 $S_q(r)$ spectra, that is, smaller $\zeta (q)$ values, for a given $q$, and 
(iii) flaring ARs tend to show more nonlinear $\zeta (q)$ curves than non-flaring 
ARs. These tendencies, nonetheless, have exceptions for both flaring and 
non-flaring ARs so a comprehensive statistical study relying on a large 
active-region sample is required. The result should be a statistical preference 
on the values of $S_q(r)$ and $\zeta (q)$ for flaring ARs which might improve 
our flare-forecasting ability. 
\item[[4]] An interesting result is found regarding the temporal evolution of 
the scaling exponent $\zeta (q)$ in a flaring AR. The values of $\zeta (q)$ appear 
to increase significantly a few $hr$ before the flare and decrease significantly 
after the event. This corresponds to a decrease of the intermittency in the 
active-region magnetic fields during the pre-flare phase followed by a subsequent 
increase of the intermittency in the post-flare phase. This result, however, is 
only based on a single example of a flaring AR with flares occurring during the 
observations, as compared to non-flaring ARs. Further study of more flaring 
ARs is obviously required. If confirmed, this trend might be of specific 
flare-predictive value. 
\een

To complete the discussion on the structure functions $S_q(r)$, we notice that 
a largely overlooked aspect of their analysis is the lower limit $r_1$ of their 
turbulent inertial range (Figure \ref{f3}c). In several examples 
(Figures \ref{f6}a, \ref{f7}a) and for moderate selectors ($q \le 2$), $r_1$ 
extends down to the instrument's pixel size. 
This feature can be useful in future, 
high-resolution, magnetograms and can be used to identify the length scale 
where the breakdown of the turbulent inertial range occurs. This is the length 
scale where magnetic resistivity sets in or, in other words, the magnetic 
Reynolds number becomes small enough to allow dissipation of free magnetic 
energy via magnetic reconnection. Finding the true value of $r_1$ will allow 
testing of theoretical estimations and their respective physical backgrounds, 
such as the Taylor microscale or Kraichnan's length scale of MHD turbulence. 
This development will hardly contribute to our space weather forecasting 
ability but it will 
advance our understanding of solar magnetic fields. The Solar-B 
mission will contribute very high-resolution space-based vector magnetograms 
in a few years. The quantitative study of turbulence and the flaring phenomenon 
in these magnetograms will certainly be a worthy task.  
\acknowledgements
I would like to thank L. Vlahos for a long, fruitful, collaboration and B. J. 
LaBonte for contributing the IVM vector magnetograms. Data used here from the 
Mees Solar Observatory, University of Hawaii, are produced with the support of 
NASA and AFRL Grants. 
%

\theendnotes

\end{article}
\end{document}